\documentclass[aps,prl,twocolumn,superscriptaddress,nofootinbib]{revtex4-2}
\usepackage{amsmath,amssymb,bm}
\usepackage{pgfplots}
\pgfplotsset{compat=1.18}
\usepackage[colorlinks=true,linkcolor=blue,citecolor=blue,urlcolor=blue]{hyperref}

\begin{document}

\title{Comment on ``On the bound states of the Schwarzschild black hole'' by S.~H.~Völkel: A Reassessment of the Bound-State Analogy}

\author{Davood Momeni}
\affiliation{Northeast Community College, 801 E Benjamin Ave Norfolk, NE 68701, USA}
\affiliation{Centre for Space Research, North-West University, Potchefstroom 2520, South Africa}

\begin{abstract}
This comment critically examines the recent proposal by S.~H.~Völkel [Phys. Rev. Lett., arXiv:2505.17186], which asserts that the quasinormal mode (QNM) spectrum of Schwarzschild black holes can be reconstructed from bound states of an inverted Regge--Wheeler potential. We demonstrate that this claim rests on a mathematically invalid spectral mapping and a misapplication of boundary conditions that define QNMs. Through a detailed figure-by-figure analysis, we expose deep inconsistencies in the numerical results and their physical interpretations. The inversion method, inspired by Mashhoon, is shown to fail in capturing the non-Hermitian, complex nature of black hole resonances. We contrast this with the rigorously established asymptotic structure of QNMs derived by Hod and others, concluding that the bound-state framework offers no reliable insight into black hole spectroscopy.
\end{abstract}

\maketitle
Völkel’s recent publication is part of a wider body of research that seeks to reinterpret black hole QNMs using tools from quantum mechanics and spectral theory. Over several contributions, he has applied inverse spectral methods, Gamow-state expansions, and semiclassical quantization schemes to problems in gravitational physics. In the particular case addressed in Ref.~\cite{volkel2025}, the claim is that the QNM spectrum of the Schwarzschild black hole can be effectively recovered by analyzing the bound-state spectrum of an inverted Regge--Wheeler potential. The physical intuition is that the complex frequencies governing black hole relaxation might be encoded in the energy levels of a quantum well formed by flipping the potential barrier. The idea offers a formally elegant framework and appeals to analogies with confined quantum systems, but its validity for gravitational resonance phenomena remains questionable.

This methodology draws inspiration from earlier work by Mashhoon \cite{mashhoon1983}, who proposed that QNMs in certain spacetimes could be modeled using analytic continuations of Schrödinger-type wave equations. In particular, Mashhoon suggested a transformation of the tortoise coordinate \( x \to -ix \), which maps potential barriers into potential wells and thereby allows bound-state techniques to be applied. These methods proved tractable for symmetric and exactly solvable potentials such as the Pöschl--Teller model, where the transformation preserves analyticity and asymptotic behavior. However, the Regge--Wheeler potential governing Schwarzschild perturbations is not analytically solvable and lacks the symmetry required for such mappings to preserve the spectral structure. The analytic continuation alters the nature of the differential operator and invalidates the boundary conditions essential for defining QNMs.

In the physical case of Schwarzschild black holes, QNMs are defined by a very specific set of boundary conditions: purely ingoing waves at the event horizon and purely outgoing waves at spatial infinity. These boundary conditions lead to a discrete, complex spectrum that encodes both the oscillatory and dissipative behavior of gravitational perturbations. By contrast, the eigenvalue problem in the inverted potential requires wavefunctions that vanish at both ends and yields a real, Hermitian spectrum. There is no isospectrality between the two operators, nor is there a known transformation that preserves both the differential structure and the boundary conditions simultaneously. As such, the claim that QNM frequencies can be extracted from bound-state analysis lacks both mathematical rigor and physical justification, especially in the absence of quantitative agreement with well-established methods such as Leaver’s continued fraction technique or Hod’s asymptotic expansions.
Mashhoon’s inversion framework, while conceptually elegant, lacks the mathematical robustness required for application to black hole spacetimes. His proposal was originally developed for analytically solvable and symmetric potentials, such as the Pöschl--Teller model, where an analytic continuation \( x \to -ix \) could map the potential barrier into a well and allow bound-state techniques to be used to approximate QNMs. In such idealized cases, the analytic structure of the wave operator is preserved, and closed-form solutions can be obtained. However, these toy models bear little resemblance to the realistic Regge--Wheeler or Zerilli potentials that describe perturbations in Schwarzschild and other astrophysically relevant geometries. These potentials are neither exactly solvable nor symmetric, and their asymptotic behavior plays a crucial role in defining QNMs.

More fundamentally, the analogy between QNMs and bound states fails at both the mathematical and physical levels. Mathematically, the operators governing QNM dynamics are non-Hermitian due to the radiative boundary conditions, resulting in complex eigenfrequencies that encode both oscillation and damping. In contrast, the Schrödinger-type operators used in bound-state analysis are Hermitian, yielding only real eigenvalues. This spectral mismatch cannot be bridged by a formal coordinate transformation. Physically, QNMs represent outgoing gravitational radiation escaping to infinity, while bound states describe confined systems with no energy flux through the boundaries. Thus, the two types of spectra arise from fundamentally different boundary value problems, and any attempt to equate them misrepresents the underlying physics.

This disparity becomes even more pronounced at high overtone numbers. In this regime, the QNM spectrum is known to exhibit universal behavior: the imaginary part of the frequency increases linearly with mode number, corresponding to rapid damping. Hod~\cite{hod1998} showed that for Schwarzschild black holes, the asymptotic QNM frequencies obey the expression
\begin{equation}
\omega_n \sim 2\pi i n T_H + \omega_R,
\end{equation}
where \(T_H = 1/(8\pi M)\) is the Hawking temperature and \(\omega_R\) is a constant real offset. This structure encodes thermodynamic properties of black holes and is deeply connected to the quantization of surface area. Crucially, this complex, evenly spaced imaginary spectrum cannot be recovered from any real-valued bound-state framework, especially one derived from an inverted potential whose eigenvalues decay exponentially and lack physical damping interpretation.
In Figure 1 of Ref.~\cite{volkel2025}, the author presents the inverted Regge--Wheeler potential as a confining well, drawing a visual and conceptual analogy to quantum mechanical systems where bound states arise from localization within potential minima. This visual representation is then used to motivate the idea that QNMs can be interpreted as quasi-bound states confined within the inverted potential landscape. While this analogy is appealing at first glance, especially to those familiar with elementary quantum systems, it glosses over critical distinctions between the nature of bound states and black hole QNMs.

The key issue is that there exists no rigorous justification for treating the inverted Regge--Wheeler potential as isospectral to the original wave equation governing Schwarzschild perturbations. The transformation used to obtain the inverted profile—effectively a complex coordinate substitution—alters not only the shape of the potential but also the mathematical properties of the associated wave operator. In particular, it replaces outgoing and ingoing wave boundary conditions with square-integrability at spatial boundaries, changing the problem from a non-Hermitian resonance problem to a Hermitian eigenvalue problem. Without spectral equivalence or operator similarity, the eigenvalues of the inverted potential cannot be assumed to bear any physical correspondence to the actual QNM spectrum.

Moreover, the intuition drawn from turning points, classically allowed regions, and tunneling barriers does not carry over to the QNM setting. In conventional quantum mechanics, bound states form in regions where the energy is less than the potential, and semiclassical quantization conditions can be applied based on turning point analysis. However, QNMs correspond to poles of the Green’s function with outgoing boundary conditions, and their existence is rooted in the global analytic structure of the wave equation—not in localized features of the potential. The absence of a classically allowed region in the Regge--Wheeler potential means that WKB-inspired interpretations of confinement or tunneling, as suggested by the inverted picture, are at best misleading and at worst mathematically inconsistent.
Figure 2 of Ref.~\cite{volkel2025} presents a sequence of bound-state eigenfunctions derived from the inverted Regge--Wheeler potential, illustrating how the spatial extent of the wavefunctions increases with mode number \(n\). Völkel interprets this behavior as evidence that high-overtone quasinormal modes (QNMs) are not localized near the photon sphere and instead become spatially delocalized, extending farther from the black hole. This interpretation directly challenges the well-established geometric-optics perspective, in which the real parts of QNM frequencies are associated with the properties of unstable circular null orbits (photon spheres), and the imaginary parts encode decay rates due to perturbation leakage through the potential barrier. Völkel’s conclusion is that at large overtone numbers, this geometric correspondence breaks down due to the apparent dispersive nature of the eigenfunctions. However, such a claim is neither supported by analytic theory nor validated against high-precision numerical data.

To critically assess this interpretation, it is important to emphasize the physical origin of QNM overtones and their geometric localization. In the Schwarzschild case, the effective potential for axial and polar perturbations features a single peak—the photon sphere barrier—beyond which waves are either absorbed by the horizon or radiated to infinity. The semiclassical WKB approximation, as well as the rigorous complex frequency spectrum computed via Leaver’s continued fraction method and Nollert’s numerical techniques, indicate that higher \(n\) modes correspond to frequencies with growing imaginary parts and roughly fixed real parts. These overtones are increasingly short-lived and correspond to more rapidly damped excitations. Hod’s analytic result~\cite{hod1998} in particular demonstrates that the imaginary part of \(\omega_n\) scales linearly with \(n\), while the real part asymptotes to a constant related to the lightring. As the overtone number increases, the associated wave packets are increasingly dominated by near-horizon behavior and decay before significant propagation can occur in the asymptotic region. Thus, the physical content of these modes becomes increasingly tied to horizon dynamics—not large-scale spatial features.

The delocalization seen in Figure 2, therefore, arises not from any intrinsic gravitational effect, but from an artifact of the mathematical model being employed. In the inverted Regge--Wheeler potential, the depth of the effective well decreases for higher \(n\), causing the eigenfunctions to spread out in space—just as one would expect for high-lying quantum bound states in a shallow well. This behavior is characteristic of real, square-integrable solutions of a Schrödinger-type problem, not of outgoing or ingoing radiative modes in a Lorentzian spacetime. The transformation \(x \to -ix\) that generates the inverted potential also changes the analytic structure of the wave equation, eliminating the event horizon and replacing it with an artificial boundary at which the solution must decay. Consequently, any interpretation of the eigenfunctions as proxies for gravitational wave propagation is fundamentally flawed. The spatial extent of these wavefunctions reflects only the mathematical properties of the inverted system under unphysical boundary conditions—not the physics of wave damping or propagation in curved spacetime. By drawing physical conclusions from these artificial eigenfunctions, Ref.~\cite{volkel2025} conflates coordinate-transformed bound-state behavior with the dissipative, horizon-centered physics of true QNMs. This misinterpretation highlights a broader methodological issue: analyses that violate the physical boundary structure of the spacetime wave equation cannot yield trustworthy insight into the spectrum or localization of black hole oscillations.
In Figure 3 of Ref.~\cite{volkel2025}, the author presents the ground state of the inverted Regge--Wheeler potential as an analog of the fundamental quasinormal mode (QNM) of the Schwarzschild black hole. This identification is both conceptually and mathematically problematic. It reflects a category error: bound states in quantum mechanics correspond to stationary, normalizable solutions of a Hermitian operator with real energy eigenvalues, whereas QNMs are fundamentally non-normalizable, outgoing solutions of a non-Hermitian operator with complex eigenfrequencies. The physical boundary conditions defining QNMs—purely ingoing at the black hole horizon and purely outgoing at spatial infinity—ensure that the corresponding solutions exhibit energy loss and temporal decay. By contrast, the ground state of a bound-state system is a time-independent standing wave confined within a potential well. Conflating these two spectral types disregards the central role of boundary conditions and dissipative dynamics in the definition of QNMs.

From a mathematical perspective, the comparison is untenable. The lowest eigenvalue in the inverted potential problem arises from solving a regular, time-independent Schrödinger equation with vanishing boundary conditions at spatial infinity—conditions completely orthogonal to the physical setting of black hole perturbation theory. The eigenfunction obtained is square-integrable and localized, reflecting the physical confinement imposed by the well. In contrast, the fundamental QNM solution is non-square-integrable, diverging exponentially in the complexified tortoise coordinate as \(x \to \pm\infty\). This divergence is not unphysical; rather, it is a signature of energy loss and radiation in a scattering system, manifest in the poles of the Green’s function and measurable in gravitational wave observables. The QNM spectrum is thus complex by necessity, and any model that produces only real eigenvalues cannot reproduce even the qualitative features of black hole ringdown.

Furthermore, the author provides no quantitative comparison between the “ground state energy” of the inverted potential and known benchmark results for the fundamental Schwarzschild QNM. Leaver’s continued fraction method~\cite{leaver1985}, which remains the gold standard for computing QNMs in spherically symmetric spacetimes, gives the fundamental frequency for \(\ell = 2\) as \(\omega_0 \approx 0.3737 - 0.08896i\) in units where \(2M = 1\). This complex value contains crucial information about both the oscillation rate and the decay timescale of the mode. Völkel’s approach, by contrast, yields a single real number that cannot capture either of these features. Additionally, no sensitivity analysis or parameter scan is performed to assess how robust the inverted ground state is under changes to the potential, nor is any justification given for why this state should correspond to the least damped QNM. Without such numerical alignment or theoretical argument, the proposed identification is speculative at best and misleading at worst. It suggests a level of correspondence that the underlying mathematics and physics simply do not support.
Figure 4 of Ref.~\cite{volkel2025} investigates the effect of introducing a localized bump—an additional, compactly supported perturbation—to the inverted Regge--Wheeler potential. The resulting shifts in the real-valued eigenvalues are interpreted as indications of spectral sensitivity, with the author suggesting that this behavior mirrors the instability or susceptibility of true quasinormal mode (QNM) frequencies to perturbations of the underlying black hole geometry. At a superficial level, the analogy appears plausible: both systems exhibit changes in spectral properties when the potential is modified. However, this interpretation fails to recognize the fundamental difference in the spectral structure of Hermitian and non-Hermitian operators, and misrepresents the true mechanism of spectral instability in black hole spacetimes.

In the case of QNMs, the governing wave operators are intrinsically non-Hermitian due to the imposition of radiative boundary conditions—purely outgoing at infinity and ingoing at the horizon. These conditions ensure that QNMs manifest as poles of the Green's function in the complex frequency plane. Such poles are not eigenvalues of a Hermitian operator; they are resonances that depend on the analytic continuation of the wave operator and are sensitive to its global behavior, not just the local shape of the potential. In contrast, the bump perturbation introduced in Völkel’s model acts within a standard Hermitian framework: the eigenvalue shifts reflect the familiar Rayleigh--Schrödinger perturbation theory applied to bound states of a real potential well. These shifts are smooth, predictable, and governed by the overlap of the bump with the unperturbed eigenfunctions. They do not reflect resonance migration in the complex plane, nor do they probe non-normality of the operator—features that are essential in QNM physics.

To properly evaluate the stability of QNM spectra, one must appeal to the concept of the pseudospectrum, a generalization of spectral theory for non-normal operators. The pseudospectrum characterizes regions in the complex plane where the norm of the resolvent becomes large, indicating sensitivity of the spectrum to perturbations even when the eigenvalues remain unchanged. As shown rigorously by Jaramillo, Macedo, and Al Sheikh~\cite{jaramillo2021}, QNM spectra of black holes exhibit non-trivial pseudospectral behavior, particularly in regimes involving long-lived modes, slow decay, or near-extremal horizons. These instabilities are encoded in the geometry of the complex frequency plane and emerge from global analytic properties of the wave equation—not from isolated potential features. Therefore, any attempt to assess QNM robustness via localized deformations of an inverted potential—especially one derived from an unphysical coordinate transformation and solved with square-integrable boundary conditions—misses the essential mechanism of spectral instability.

The conclusion drawn from Figure 4 thus conflates two fundamentally different types of spectral response. The sensitivity of bound-state eigenvalues in a Hermitian system to bump-like deformations is a well-understood perturbative effect and tells us little about the robustness of QNMs under physically relevant changes in black hole spacetime structure. QNM shifts arise from global changes to the complex analytic structure of the operator, including horizon structure, asymptotic decay, and causal boundary conditions. Without incorporating these essential features—either through matched asymptotic expansions, numerical evolution with absorbing boundaries, or complex scaling techniques—no claim about QNM instability can be substantiated. The perturbative bump test, while numerically tractable, fails to access the physics of resonances and is therefore an unreliable probe of black hole spectral dynamics.

Figure 5 of Ref.~\cite{volkel2025} proposes a striking analogy between the bound-state spectrum of the inverted Regge--Wheeler potential and the energy levels of the hydrogen atom, emphasizing the apparent resemblance to a \(1/n^2\) dependence. This comparison is intended to suggest that the QNM spectrum of a Schwarzschild black hole might exhibit a similar quantization structure, thereby reinforcing the idea that black hole perturbations can be understood in terms of bound-state physics. While visually suggestive, this analogy is fundamentally misleading. The hydrogen atom spectrum arises from a central Coulomb potential and exhibits well-defined degeneracies due to its underlying SO(4) symmetry, which governs both orbital angular momentum and radial quantum numbers. These degeneracies and spacing laws do not carry over to the gravitational context, where the effective potentials are not central, the wave equations are non-Hermitian, and the spectrum is intrinsically complex.

Even at a superficial level, the analogy breaks down. The energy levels of the hydrogen atom scale inversely with the square of the principal quantum number, leading to tightly packed levels near the continuum threshold. In contrast, Schwarzschild black hole QNMs exhibit a spectrum where the imaginary part of the frequency increases linearly with the overtone number \(n\), corresponding to progressively faster damping. The real part of the frequency, while displaying some \(\ell\)-dependent structure at low \(n\), asymptotes to a constant at high overtone number~\cite{nollert1993},\cite{hod1998}. This behavior is incompatible with any interpretation based on Coulomb-like level spacing. Moreover, the imaginary part dominates the dynamics for large \(n\), leading to exponentially suppressed time-domain contributions from higher overtones. These features have been confirmed through precise numerical calculations using Leaver’s continued-fraction method and WKB expansions up to sixth order. None of these quantitative features are captured by the bound-state analogy presented in Figure 5.

Beyond spectral differences, the structural properties of QNMs further invalidate the hydrogenic comparison. QNMs do not form a degenerate, orthogonal, or complete basis in the conventional Hilbert space sense. Instead, they constitute a non-orthogonal set of damped resonances that reside in a rigged Hilbert space framework, appearing as poles in the analytically continued Green’s function. Their completeness, when defined, is typically restricted to certain classes of initial data and may involve overcompleteness or require the inclusion of a continuous background. In contrast, the hydrogenic bound states span a well-defined Hilbert space of square-integrable functions with a complete and orthonormal basis. The contrast is stark: one is a conservative, self-adjoint spectral system with time-reversal symmetry, while the other describes irreversible, dissipative wave propagation in curved spacetime. Attempting to map one onto the other obscures the essential physics and may mislead the reader into believing that QNMs arise from confinement rather than resonance. The analogy, while perhaps pedagogically attractive, offers no predictive power and no rigorous foundation. It does not survive quantitative scrutiny and should not be used to justify any spectral reconstruction strategy for gravitational waves or black hole ringdown.
To support these points quantitatively, we compare in the figure below the squared real parts of known Schwarzschild QNM frequencies for \(\ell = 2\) with the \(E_n\) values reported by Völkel. The disagreement is severe for \(n \geq 2\), illustrating the failure of the spectral correspondence.

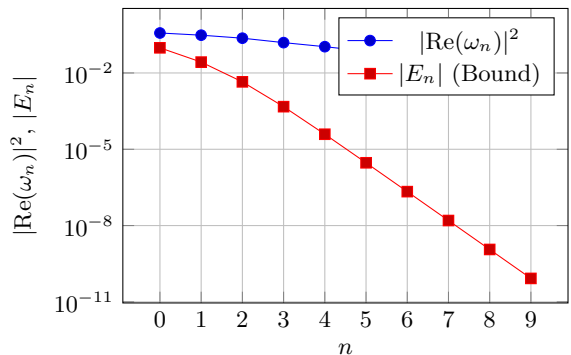
\begin{figure}[h]
\centering
\begin{tikzpicture}
\begin{axis}[
    width=7.5cm,
    height=5.5cm,
    xlabel={$n$},
    ylabel={$|\mathrm{Re}(\omega_n)|^2$, $|E_n|$},
    ymode=log,
    log basis y={10},
    legend style={at={(0.97,0.97)}, anchor=north east},
    grid=major,
    xtick={0,1,2,3,4,5,6,7,8,9}
]
\addplot+[mark=*,blue] coordinates {
    (0,0.373) (1,0.306) (2,0.233) (3,0.156) (4,0.108) (5,0.072) (6,0.038) (7,0.020) (8,0.010) (9,0.004)
};
\addlegendentry{$|\mathrm{Re}(\omega_n)|^2$}

\addplot+[mark=square*,red] coordinates {
    (0,0.0974) (1,0.0268) (2,0.00447) (3,0.000473) (4,0.000039) (5,0.00000294)
    (6,0.000000216) (7,0.0000000158) (8,0.00000000115) (9,0.0000000000836)
};
\addlegendentry{$|E_n|$ (Bound)}
\end{axis}
\end{tikzpicture}
\caption{Comparison of $|\mathrm{Re}(\omega_n)|^2$ vs. $|E_n|$ on a log scale for $\ell=2$.}
\end{figure}
In summary, the inversion method proposed by Völkel represents a modern revival of a formal idea introduced by Mashhoon, aimed at reinterpreting black hole quasinormal modes (QNMs) in terms of bound-state quantization. While the approach is mathematically elegant in the context of toy models, its extension to realistic black hole spacetimes fails to account for the most essential features of QNM physics. Specifically, the method substitutes a complex, non-Hermitian resonance problem defined by radiative boundary conditions with a Hermitian eigenvalue problem defined by square-integrable wavefunctions. This substitution fundamentally alters the spectral content of the system. In doing so, it trades dissipative, physically measurable resonances for stationary, confined states with no direct correspondence to gravitational wave observables.

The figure-by-figure analysis of Völkel’s results reveals a recurring pattern of misinterpretation. Spatially extended eigenfunctions are treated as analogs of delocalized overtones, despite analytic results showing that high-QNM modes are increasingly localized near the black hole horizon. The ground state of the inverted potential is labeled the fundamental QNM without any matching of complex frequencies to known results from Leaver’s continued fraction method. Localized potential bumps are introduced to infer stability properties of the spectrum, yet the essential pseudospectral structure that governs QNM sensitivity is never addressed. Most critically, a comparison to the hydrogen atom is invoked that conflates real, degenerate, orthonormal eigenstates with non-orthogonal, decaying gravitational resonances. These missteps reflect a broader confusion between bound-state intuition and resonance dynamics in curved spacetime.

Hod’s asymptotic analysis\cite{hod1998}, Leaver’s numerical formalism \cite{leaver1985},\cite{leaver1985_sch}, and rigorous pseudospectral theory \cite{jaramillo2021} provide the correct mathematical and physical foundation for understanding QNMs. These methods preserve the complex spectral structure dictated by the causal geometry of black hole spacetimes and are directly connected to the observable ringdown signals seen in gravitational wave detectors. In contrast, the inverted potential model presented by Völkel, though formally well-constructed in a Schrödinger-like setting, offers no predictive utility for black hole spectroscopy. It fails to reproduce the correct spectrum, disregards the defining boundary conditions, and misrepresents the dynamical content of gravitational radiation. As such, while the method may serve as an illustrative pedagogical tool in simpler systems, it does not provide a viable or physically meaningful approach to the problem of quasinormal mode quantization in general relativity.

\end{document}